\documentclass[11pt,twoside]{article}

\usepackage{asp2006}
\usepackage{epsf}
\usepackage{lscape}
\usepackage{natbib}
\usepackage{url}
\usepackage{graphicx}

\begin{document}
\title{Summary of Panel Discussion II: Theoretical Models
and Observational Constraints in High-Mass Star Formation}   
\author{H.~Zinnecker}   
\affil{Astrophysical Institute Potsdam, An der Sternwarte 16, 14482 Potsdam, Germany}
\author{H.~Beuther}
\affil{Max Planck Institute for Astronomy, K\"onigstuhl 17, D-69117 Heidelberg, Germany}    

\begin{abstract}
  This paper presents a summary of a panel discussion to more directly
  confront theoretical models and observational constraints in massive
  star formation research.  The panel was moderated by Hans Zinnecker,
  and panel members included Ian Bonnell, Chris McKee, Francesco
  Palla, Malcolm Walmsley, and Harold Yorke.  Additional ample
  discussion with the audience is recorded.
\end{abstract}
\label{panel2}


\section{Concept and Introduction}

This is a summary of the second panel discussion at MSF07 conference
in Heidelberg on Sept 13, 2007, entitled: Theoretical Models and
Observational Constraints, which lasted about 90 min. Hans Zinnecker
was the moderator of this discussion. The invited participants of the
panel included the following five distinguished massive star experts:
Ian Bonnell, Chris McKee, Francesco Palla, Malcolm Walmsley, and
Harold Yorke (almost all theorists).  The idea was to first give the
word to the panelists who would each raise and discuss the one
observational question which they found most interesting and
important.  In the second part the audience, now stimulated, would be
invited to join the discussion and ask their ``tough'' theory
questions to the panelists or suggest further observational homework
to us all.  This more or less worked, and people can refresh their
memory of what has been said via the video of the panel discussion
available at \url{http://www.mpia-hd.mpg.de/MSF07/}.

At the very beginning the moderator took a few minutes to introduce
the subject by displaying the three basic models of massive star
formation: monolithic collapse, competitive accretion, and stellar
mergers (cf. \citealt{zinnecker2007a,beuther2007}). He then asked the
question which observational constraints could be placed on the
various models and whether massive star formation is or is not a
scaled-up version of low-mass star formation.  He surmised that all
the models could be mutually correct and applied to different
circumstances of massive star formation. Thus, the analogy is that
three theoretical models set forth could be imagined as the two sides
of the same coin, plus its rim.

\section{Key questions from the panelists}

The moderator suggested Malcolm Walmsley to go first, followed by
Chris McKee, Ian Bonnell, and Harold Yorke (in that order), and
Francesco Palla at the end.

{\bf Walmsley}, after admitting that he is still confused, chose his
outstanding question to be the following: {\it What are the accretion
  rates onto massive protostars, and how can we best imagine to get an
  estimate?} Three types of accretion flows must be distinguished,
that from the disk onto the star (not yet observed), that through the
disk (enabling gravitational torques), and that onto the disk from a
protostellar envelope. In a time-averaged sense, all these mass
accretion rates should be the same.  He referred to the talks of Yorke
and Hosokawa whose results showed that different amounts of accretion
onto the central star make a big difference in the stellar evolution
(see also their contributions to this volume).

{\bf McKee}, addressing his question to the observers in the audience,
asked: {\it Is turbulence universal and what drives it?} He referred
to his own turbulent core model (i.e., monolithic collapse model) for
massive star formation as one in which turbulence is driven, while the
competitive accretion model is characterized by an initially high
level of turbulence which then decays. He referred to the review of
\citet{maclow2004} for a discussion what can be the drivers of
turbulence.  As time goes on, using increasingly sophisticated
computer codes we might come to a concensus. While we know that there
is more turbulence on larger scales (e.g., the mm observations of
Chris Brunt presented at this meeting), he emphasized that we need to
know much more about the smaller scales. He challenged the observers
to find out which kind of turbulence is more typical: driven or
decaying?

{\bf Bonnell}, after starting with a funny analogy for competitive
accretion (go to a busy restaurant instead of an empty one!), raised
the key issue: {\it Is all massive star formation clustered?} In the
competitive accretion model, all massive star formation is linked to
cluster formation, and the answer to the question would be YES.  He
suggested observations to look for clustering statistics around stars
of (say) greater than 10\,M$_{\odot}$, and referred to the work of
\citet{weidner2006} on the maximum stellar mass as a function of the
cluster mass, but also pointed out that Elmegreen disagrees with this
work.

{\bf Yorke}, after endorsing Walmsley's question as a primary question
that he too would have posed, found himself left with a second
important question, viz. {\it Is the upper IMF and the maximum stellar
  mass universal?} He wondered why the upper IMF in the Orion Nebula
Cluster and in the R136 cluster in the LMC could be the same. He
emphasized that very high angular resolution is needed to be sure we
are observing stars and not whole unresolved star clusters.  He also
wondered whether a 1:1 mapping of the molecular core mass function,
which looks similar as the IMF, can really be true, given the very
high multiplicity fraction of massive stars.

{\bf Palla}, alluding to homework for the next conference on massive
stars, presented us and himself with the following final panelist
question: {\it Is high-mass star formation first or last?} In other
words, he asked: what is the sequence of events, e.g., in the making
of a star cluster, and are there ways to discriminate the order in
which low-mass stars and high-mass stars form? He emphasized two
aspects: (1) what are the accretion rates for low-mass and
high-mass stars? (2) It seems clear that often they form together in
the same environment (clusters).

The moderator used his prerogative to interfere and proposed to have a
poll at this point to the panelists requesting to answer Palla's
question by a show of hands.  Malcolm was in favor of ``last'', Chris
and Ian (miraculously) agreed ``at the same time'', Harold also held
``at the same time'' while Palla voted for ``last'' - as did the
moderator, who added his comment that if massive stars formed first
(which none actually claimed) we would have a serious problem with
radiative feedback which would destroy the parent cluster cloud and
prevent low-mass star formation.

\section{Additional questions from the audience}

In the following, we summarize the questions and comments received
from several members of the audience, including reactions from the
panelists and the moderator.\\

\noindent {\bf Tan:} I'd like to take issue with Ian Bonnell who said
that the turbulent core model and the competitive accretion model are
not so different after all.  I think there is a big difference. In the
turbulent core model the gas gathers first. My question to the
observers would be: can we try to catch protostars with (or without)
gas mass around them and find out whether such gas is bound to the star?\\

\noindent {\bf Churchwell:} 1) How does matter get onto the central star? We
don't know.  2) How can massive star formation proceed in isolation?
How to prevent fragmentation?

\noindent {\bf Yorke reply:} ad 1) We don't know for low-mass stars either, nor for AGN.

\noindent {\bf Palla reply:} ad 2) With new survey data (like Spitzer), we will
be in a better position to estimate the fraction of isolated massive
stars.

\noindent {\bf McKee reply} ad 2) Some massive O-stars are found in small
groups. There seem to be low-mass clouds producing untypical IMFs -
Krumholz and McKee are exploring this scenario, where a high threshold
pressure rules.  Fragmentation is prevented when the accretion rate is
very high, causing gas heating.
       
\noindent {\bf Bonnell reply:} ad 2) Even with heating (later on), initially
there is no heating to prevent fragmentation.\\

\noindent {\bf Shepherd:} I think we have to iterate between observations
and theory to get the correct answers. For example, what
is the outflow rate and the turbulent feedback from massive stars?
Can theoretical simulations help us to correct the observations?

\noindent {\bf Yorke reply:} maybe in 4 years but not in the immediate
future...

\noindent {\bf McKee comment:}
We also need surveys as to the generality of turbulence.\\

\noindent {\bf Krumholz:} I believe where we did make progress in the
last few years is whether star formation is slow or fast. Observers
can go out and test how the rate of star formation depends on gas
density and environmental factors.  The rate of star formation has
implications on the mode of star formation \citep{krumholz2007}.  It
is about 1\% of the cloud mass per free-fall time, thus it is slow...
there is no evidence that it would accelerate with time...

\noindent {\bf Bonnell comment:} ... too bad Bruce Elmegreen is not here ...

\noindent {\bf Palla comment:} ... I am Bruce Elmegreen! Bruce did not
say that star formation is short-lived, he said two modes coexist in
giant molecular clouds (GMCs): one is on the dynamical timescale, but
the way it gets there is slow (i.e., there is another, longer
timescale due to the fact that GMCs have envelopes which are less
dense).  Now I am Palla: You also have to study the stars (not only
the gas) to find the star formation rate, particularly in clusters!

\noindent {\bf Henning:} Bruce Elmegreen was here two weeks ago at a
summer school, he should have stayed on. We discussed the IMF of cores
and the IMF of stars. We noted that the core IMF stops at 8-10
M$_{\odot}$, i.e., massive cores are rare or don't exist. The maximum
core mass known is 23 M$_{\odot}$, how do you form a 50 M$_{\odot}$
star from it?

  \noindent {\bf McKee reply:} It is true that we have not found the
  high-mass cores yet. The prediction, however, is that ALMA will find
  hundreds of massive cores in the future at large distances.  It'll
  also unravel the structure of GMCs...

  \noindent {\bf Walmsley reply:} The core mass function is a tricky
  business, as a core has no clear edge; the density profiles are such
  that most of the mass is near the edge. We need a better definition
  of "massive cores".

  \noindent {\bf Bonnell comment:} Charlie Lada's work on the Pipe Nebula seems
  to suggest that low-mass cores are unbound, while high-mass cores
  are bound... \citep{lada2008}.\\

  \noindent {\bf Vazquez:} Not only cores are not well defined, but clouds as
  well ... they are shredded and have no clear boundaries. -- I also
  have a question to the observers: can we see large-scale organized
  flows? (cf. \citealt{hartmann2007}'s model for the collapse of the
  entire Orion cloud). Can we discriminate purely random turbulence
  from focused turbulent flows?  It appears that driven and decaying
  turbulence are not so different in large-scale organized flows.\\

  \noindent {\bf Burton:} ... on the core mass function ... a warning to the
  theorists: different search algorithms give different core mass
  functions. We really have no idea!

  \noindent {\bf Zinnecker:} One should give the same set of data to two
  independent groups and perform blind tests.\\

  \noindent {\bf Bally:} Changing the subject ... We heard about the
  Arches, the Quintuplet, and the Westerlund 1 cluster which hosts a
  magnetar (black hole). Can we produce black holes by runaway mergers
  as modeled by \citet{portegies2004}?

  \noindent {\bf Zinnecker comment:} Runaway mergers are hard to
  achieve in even the densest observed clusters... but who knows maybe
  it is possible in galactic nuclei...

  \noindent {\bf Bally (continued):} What are the precursors to globular
  clusters, why don't we see them in the Milky Way?

  \noindent {\bf McKee reply:} In the Antennae, the pressure is much
  higher: you have a million solar masses squeezed into 1 pc in size.

  \noindent {\bf Yorke comment:} In the Antennae, yes; but you also have M82 or
  30 Dor. It is not an easy question to answer, it is not just
  pressure.

  \noindent {\bf Zinnecker continued:} The initial conditions for
  forming globular clusters must include lots of gas swept together
  quickly...

  \noindent {\bf McKee comment:} Here is another theoretical
  challenge: globular clusters have virtually the same stellar IMF and
  characteristic mass, despite very different conditions from
  present-day star formation (e.g., metallicity).

  \noindent {\bf Schulz:} The point was raised whether high-mass star formation
  was a scaled-up version of low-mass star formation. What about
  intermediate mass stars?

  \noindent {\bf McKee reply:} Our turbulent core model makes a link
  between low-mass and high-mass star formation: the thermal sound
  speed must simply be replaced by the turbulent sound speed for the
  higher masses.  As for the issue of outflows: the same MHD theory
  applies for low-mass and high-mass stars, which makes unique
  testable predictions for hypercompact H{\sc ii} regions (i.e.,
  outflow-confined bubbles).

  \noindent {\bf Yorke comment:} There are similarities but also differences.
  It is significant that there is no break or knee in the IMF but
  rather a gradual smooth change-over from high-mass to
  intermediate-mass and on to low-mass stars. There is one big
  difference: intermediate-mass stars end up as white dwarfs,
  high-mass stars as supernovae.

  \noindent {\bf Palla comment:} Not sure about the IMF at the highest
  masses: there could a break at (say) 30--50\,M$_{\odot}$.  The poor
  sampling at the high-mass end does not allow us to make strong
  statements.

  \noindent {\bf Zinnecker comment:} If stellar collisions start to play a role
  for the highest masses (e.g., the merging of tight massive binaries),
  then the upper IMF could have a gap before the maximum mass (the
  merger product) is reached.\\

  \noindent {\bf Hoare:} Why was triggered star formation so low on
  the list of topics? There is ample evidence from multi-wavelengths
  galactic surveys that young massive star formation is triggered. I
  want to make a plea to return to study triggered star formation. We
  are planning an IAU Symposium in Rio 2009 on this.
 
  \noindent {\bf Zinnecker comment:} There was an IAU Symposium dedicated to
  this subject at the General Assembly in Prague in 2006. If I
  remember correctly, there was no concensus on the issue ...  the
  problem is to prove a spatial and temporal causal relationship
  between the triggering agent and the triggered result. In the case
  of the Upper Sco OB association, the evidence is compelling, but in
  other cases it is circumstantial (cf. \citealt{preibisch2007}).

\section{Concluding Remarks on Methodology}

{\it Karl Popper's falsification principle:} In his concluding
remarks, the moderator wanted to leave the audience with a
philosophical thought, challenging the methodology of our daily
scientific work where we tend to find the observational evidence that
we look for to corroborate our favorite theory. It would be more
convincing to find evidence contrary to what we set out to look for.
He quoted Sir Karl Popper (1902-1994) who in 1934 in his book ``Logik
der Forschung'' suggested that a theory can never be verified or proven
but can only be falsified or disproven.  What does this mean for the
progress in our field of science?  It means that progress comes from
an iterative process, where observations not only confront theory but
also theory predicts observations. The latter can then be critically
checked by new key measurements (experimentum crucis) and can be used
to select an improved but still incomplete theory (there is no final
theory).  Actually, it would be even better if we subjected ourselves
to an even more self-critical methodology where observations confront
observations (for example at different wavelengths) and theory
confronts theory (for example simulations with different numerical
codes). Hopefully these cautious words will keep us on track in the
frantic race for new results!

\noindent \includegraphics[width=5.25in]{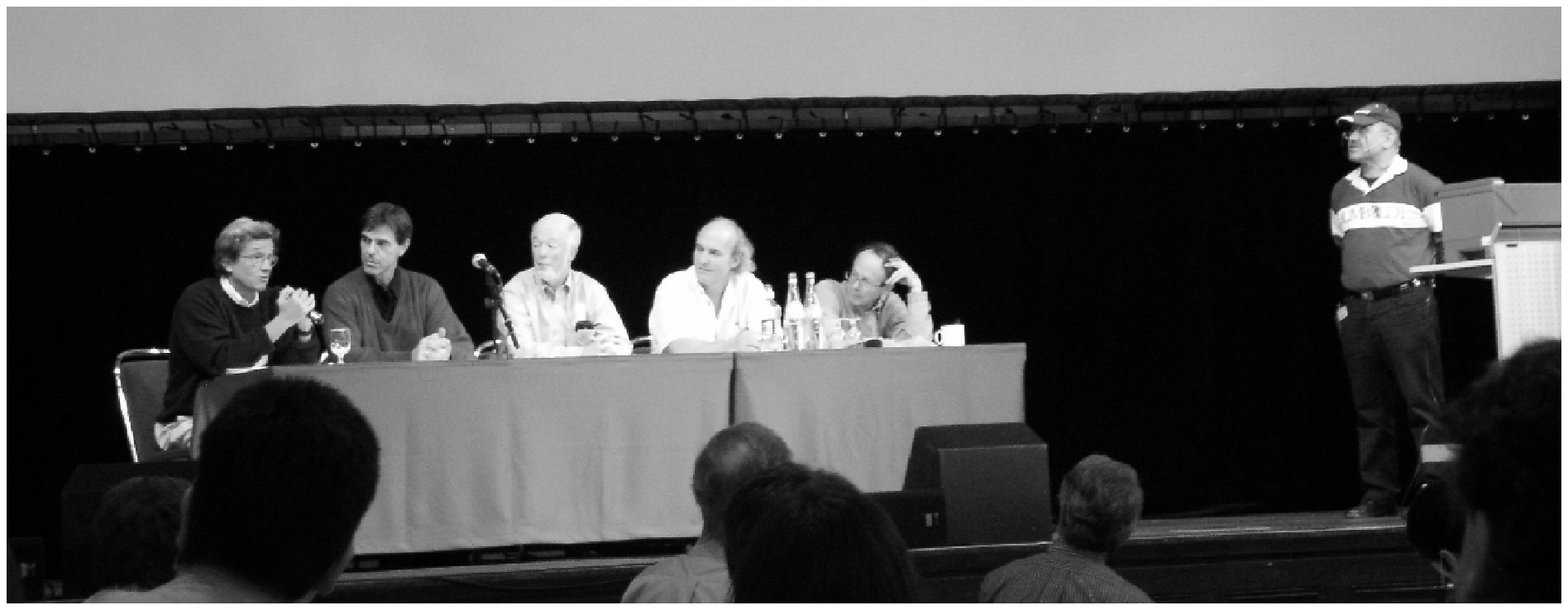}
\centerline{The panel of the second panel discussion.}

\end{document}